\documentclass[preprint,pre,aps,amsfonts,amsmath,superscriptaddress,nofootinbib]{revtex4}
\usepackage{amsmath,amssymb,amscd}
\usepackage{empheq}
\usepackage{graphicx}
\usepackage{bm}
\usepackage{color}

\newcommand{\bea}{\begin{eqnarray}}
\newcommand{\eea}{\end{eqnarray}}

\def\XXint#1#2#3{{\setbox0=\hbox{$#1{#2#3}{\int}$}
     \vcenter{\hbox{$#2#3$}}\kern-.5\wd0}}

\begin{document}
%
%%%%%%%%%%%%%%%%%%%%
%
\title{Record statistics for random walks and L\'evy flights with resetting}
\author{Satya N. Majumdar}
\email{satya.majumdar@lptms.u-psud.fr}
\affiliation{LPTMS, CNRS, Univ. Paris-Sud, Universit\'e Paris-Saclay, 91405 Orsay, France}
\author{Philippe Mounaix}
\email{philippe.mounaix@polytechnique.edu}
\affiliation{CPHT, CNRS, Ecole
Polytechnique, IP Paris, F-91128 Palaiseau, France}
\author{Sanjib Sabhapandit}
\email{sanjib.sabhapandit@gmail.com}
\affiliation{Raman Research Institute, Bangalore 560080, India}
\author{Gr\'egory Schehr}
\email{gregory.schehr@lptms.u-psud.fr}
\affiliation{Sorbonne Universit\'e, Laboratoire de Physique Th\'eorique et Hautes Energies, CNRS, UMR 7589, 4 Place Jussieu, 75252 Paris Cedex 05, France}
\date{\today}
\begin{abstract}
We compute exactly the mean number of records $\langle R_N \rangle$ for a time-series of size $N$ whose entries represent the positions of a
discrete time random walker on the line with resetting. At each time step, the walker jumps by a length $\eta$ drawn independently from a symmetric
and continuous distribution $f(\eta)$ with probability $1-r$ (with $0\leq r < 1$) and with the complementary probability $r$ 
it resets to its starting point $x=0$. This is an exactly solvable example of a weakly correlated time-series
that interpolates between a strongly correlated random walk series (for $r=0$) and an uncorrelated  time-series (for $(1-r) \ll 1$). Remarkably, we found that for every fixed 
$r \in [0,1[$ and any $N$, the mean number of records $\langle R_N \rangle$ is completely universal, i.e., independent of the jump distribution $f(\eta)$. In particular, for large $N$, we show that  $\langle R_N \rangle$ grows very slowly with increasing $N$ as $\langle R_N \rangle \approx (1/\sqrt{r})\, \ln N$ for $0<r <1$. We also computed the exact universal crossover scaling functions for $\langle R_N \rangle$ in the two limits $r \to 0$ and $r \to 1$. Our analytical predictions are in excellent agreement with numerical simulations. 
\end{abstract}
%\pacs{02.50.Ga, 05.40.Fb, 05.45.Tp}
%
\maketitle
%
%%%%%%%%%%%%%%%%%%%%
%
\section{Introduction}\label{intro}

The study of record statistics in a time series has generated a lot of interest across disciplines. For instance, this time series may represent the stock
price in finance or the daily temperature in a given city. Examples of such time series arise naturally in sports, climate data, finance, disordered systems, earthquake models etc.,  
where record statistics provide a useful tool \cite{Cha1952,FS54,hoyt,basset,SZ1999,benestad,RP2006,K2007,WK2010,AB2010,WHK2013,records_finance,WBK2011,SL2014,records_hydrology,Gembris,sports,FWK2012,GL2008,sibani,MBK2019,MPS2020} to analyse the extremes and rare fluctuations present in the data -- see Refs.~\cite{review_wergen,review_records} for recent reviews on record
statistics. Our starting point is a generic time series with $N$ entries $\{x_1, x_2, \cdots x_N\}$. A record occurs at step $m$ if $x_m > \{x_1, x_2, \cdots, x_{m-1} \}$, i.e., 
the $m$~-~th entry is bigger than all the previous entries. How many records $R_N$ occur in this time series of size $N$? It is an easily mesurable observable, that
carries interesting informations about the correlations structure present in the time series. When the
entries are random variables (either independent or correlated), $R_N$ is also a random variable and
studying its statistics is what is called ``record statistics''.  The study of record statistics started in the statistics and probability literature \cite{Res1987,ABN1992,BG2001,Nevzorov,Feller}, but off late it has generated 
much interest in the statistical physics literature also, in various contexts such as in biological evolution models \cite{KJ2005,FKAK2011,PSNK2015,PNK2016,PK2016}, random walks \cite{MZ2008,PLDW2009,Satya2010,Sanjib2011,MSW2012,WMS2012,EKMB2013,GMS2015b,GMS2016,Cha15b,MMS2020,MDMS2020a, MDMS2020b, LM2020,Kea2020,MMS2021,GL2021}, avalanches in disordered systems \cite{sibani,PLDW2009,MMR2021} etc. 
From the statistical physics point of view, one of the central issues is the universality of the statistics of $R_N$ with respect to the underlying 
probability distribution of the entries~\cite{review_records}.

The simplest and the most well studied example corresponds to the case where the entries are uncorrelated and each drawn independently from the same
parent distribution $p(x)$ -- usually referred to as the independent and identically distributed (IID) model \cite{Res1987,ABN1992,BG2001,Nevzorov}. 
In this IID case, the statistics of $R_N$ is completely universal for all $N$ as long as $p(x)$ is continuous. For example, the mean number of records in an IID time 
series of $N$ entries is given by the universal formula (see for example the review \cite{review_records})
\bea \label{av_RN}
\langle R_N \rangle = 1 + \frac{1}{2} + \cdots + \frac{1}{N} \;.
\eea
In particular, for large $N$, the mean number of records grows very slowly as 
\bea\label{av_RN_IID_as}
\langle R_N \rangle \approx \ln N \quad, \quad {\rm as} \; N \to \infty \;.
\eea
This is because it gets harder and harder to break a record as $N$
increases. 

A completely contrasting case with {\it strongly correlated} entries corresponds to the random walk model (RW) where the entry $x_n$ corresponds to the position
of a random walk (in discrete time and continuous space) after $n$ steps starting from $x_0=0$. The position of the random walker evolves via the Markov rule
\bea \label{def_RW}
x_n = x_{n-1} + \eta_n \;,
\eea
where the jump lengths $\eta_n$'s are IID, each drawn from a symmetric and continuous distribution $f(\eta)$. Note that this model also includes
L\'evy flights where the jump distribution has a fat tail: $f(\eta) \sim |\eta|^{-1-\mu}$ for large $|\eta|$ and the L\'evy index $0<\mu \leq 2$. Here, unlike the IID model,
the entries/positions $x_n  = \sum_{k=1}^n \eta_k$ for different $n$ are strongly correlated, even though the increments $\eta_k$'s are uncorrelated. In this case, the 
record statistics can be computed exactly and, interestingly, the statistics of $R_N$ is again completely universal for all $N$, i.e., independent of the jump distribution $f(\eta)$, as
long as $f(\eta)$ is symmetric and continuous. The origin of the universality in the RW model turns out to be  
very different from that in the IID model \cite{MZ2008,review_records}. In the RW model, the mean number of records is given by the universal formula \cite{MZ2008}
\bea \label{av_RN_rw}
\langle R_N \rangle = (2N+1)\,{2N \choose N}\, 2^{-2N} - 1 \;,
\eea
where, by convention, the first entry $x_0=0$ is {\it not} counted as a record. For large $N$, the mean number of records $\langle R_N \rangle$ grows much faster
than in the IID model, namely, 
\bea\label{av_RN_rw_as}
\langle R_N \rangle \simeq \frac{2}{\sqrt{\pi}}\,\sqrt{{N}} \quad, \quad {\rm as} \; N \to \infty \;.
\eea

It is natural then to ask what happens to the record statistics for a time series whose entries are neither IID nor strongly correlated
as in the RW case, in other words when the entries are only ``weakly correlated''. An ideal candidate for a model where this question
can be addressed is in fact the random walk with resetting \cite{EM1,EM2}. Random walks (or Brownian motion (BM) in continuous time) with resetting have been 
extensively studied in the recent past \cite{EM1,EM2,MV13,EM14,KMSS14,MSS15a,MSS15b,Pal15,KGN15,PKE16,NG16,Reuveni16,PR17,CS18,EM18,MPCM19a,MMSS21} -- see Ref. \cite{reset_review} for a recent review. There are two aspects of BM with resetting dynamics that have attracted much attention. (i) The first interesting fact
is that, in the presence of a resetting protocol, where a diffusing particle goes back to its initial position at
a constant rate, the dynamics violates the detailed balance and the position distribution of the
particle reaches a non-equilibrium stationary state (NESS) at long times \cite{EM1,EM2}. The relaxation dynamics to this stationary state turns out to be rather
unusual \cite{MSS15a}. Furthermore, since the Brownian motion with resetting reaches a stationary state at long times, the two-time correlation 
function between positions $\langle x(t_1) x(t_2)\rangle$ decays exponentially with the time difference $|t_1-t_2|$ \cite{MO18,SSKP21}. These ``weak correlations''
differ strongly from the standard BM without resetting where the two-time correlation function $\langle x(t_1) x(t_2)\rangle = 2 D \min(t_1,t_2)$ (where $D$ is
the diffusion constant) does not decay and $x(t_1)$ and $x(t_2)$ remain strongly correlated even when $t_2 \gg t_1$. Thus the position $x(t)$ as
a function of time in the BM with a nonzero resetting is an ideal candidate for a time series with weak correlations.  (ii) The second aspect is that a finite resetting rate 
renders the mean first-passage time (MFPT) of the Brownian particle to a fixed target finite -- in sharp contrast
to the BM without resetting where the MFPT is infinite. Moreover, there is usually an optimal 
resetting rate that minimises the MFPT, a fact that has now been verified in experiments on optical traps~\cite{BBPMC20,TPSRR20,FBPCM21}
. The existence of an optimal 
resetting rate makes this model quite attractive in the context of random search strategies \cite{EM1,EM2,EMM13,KMSS14,KGN15,CS15,PKE16,Reuveni16,BBR16,BRR20,Bres2020, Bres2021}. 

\begin{figure}[t]
\includegraphics[width = 0.8\linewidth]{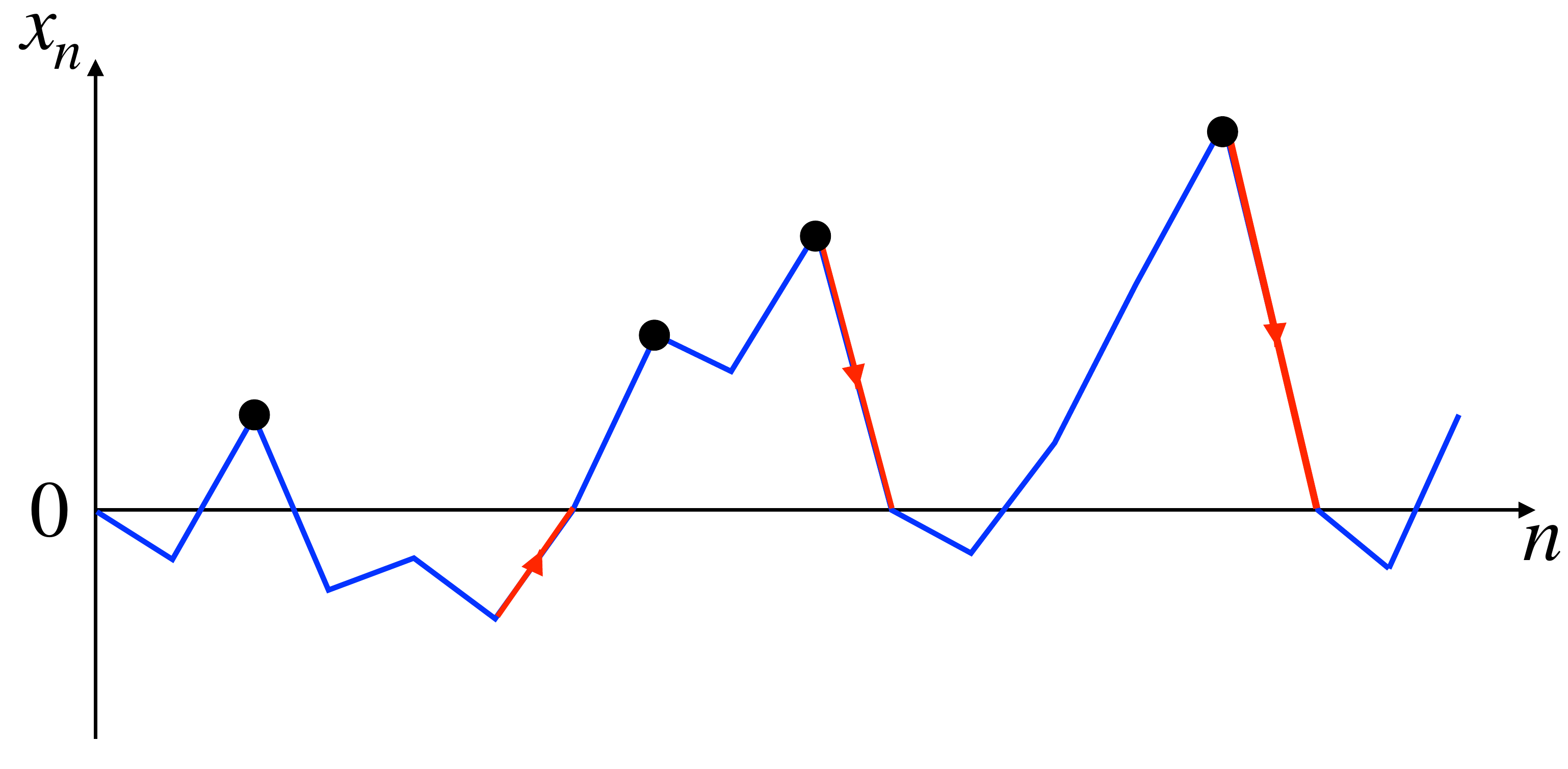}
\caption{A schematic trajectory of a random walk with resetting. The walker starts at the origin and evolves by Eq. (\ref{rw_w_reset}). With probability $r$, it resets to the origin, shown by the red arrows. A position $x_k$ at step $k$ is a record if $x_k$ exceeds all previous values. The record values in this trajectory are marked by black dots.}\label{Fig_record_RW}
\end{figure}
In this paper, 
our goal is to show that a discrete-time version of the BM with resetting \cite{KMSS14} can also be used successfully to explore other aspects going beyond these two issues (i) and (ii) 
mentioned above, namely to understand the record statistics in a weakly correlated time series. Here, the entries of our time series $\{x_0=0,x_1, x_2, \cdots, x_N\}$ 
correspond to the position of a random walker on the line in the presence of a constant resetting probability $0\leq r < 1$ to the origin. The position $x_n$ now evolves 
by the stochastic rule 
\bea \label{rw_w_reset}
x_n = 
\begin{cases}
&0 \;, \; \quad\quad\quad\quad{\rm with \; proba.\;} \, r \\
&x_{n-1} + \eta_n \;, \; \,{\rm with \; proba.\;} \, 1-r \;,
\end{cases}
\eea
where the jump length $\eta_n$ is drawn independently at each step from a continuous and symmetric distribution $f(\eta)$. In Fig. \ref{Fig_record_RW} we show schematically a typical trajectory of a random walk with resetting, where the record values are marked by black dots.

\begin{figure}[t]
\centering
\includegraphics[width = 0.7\linewidth]{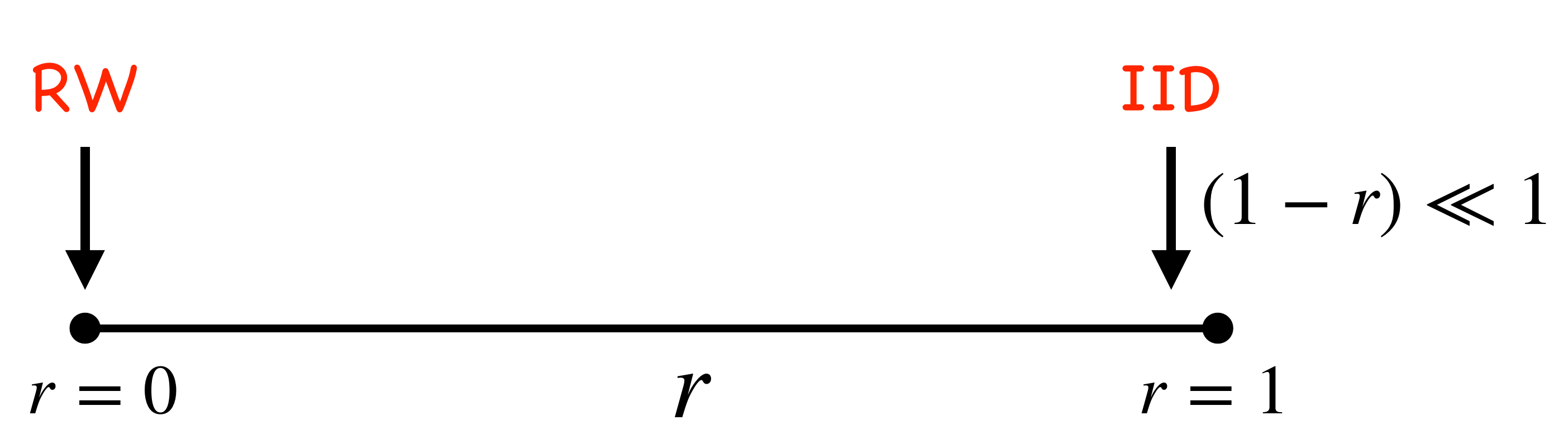}
\caption{The resetting random walker model studied here (\ref{rw_w_reset}) has a single parameter, namely the resetting probability $0\leq r < 1$. 
This model reduces to the standard random walk (marked RW) in the $r  = 0$ limit, while it corresponds to the IID model in the opposite limit $(1-r)\ll 1$.}\label{Fig_RG_flow}
\end{figure}

For $r = 0$, this model (\ref{rw_w_reset}) 
reduces precisely to a standard RW/L\'evy flight without resetting~(\ref{def_RW}). In the opposite limit $(1-r)\ll 1$, where the particle is reset to the origin with a high probability at 
all steps, only few entries, separated by an exponentially long time interval (distributed as $r^n$ where $n$ is the time interval), would be nonzero. 
Since only these nonzero entries are relevant for the record statistics (clearly an entry with value $0$ can not be a record), they are 
effectively uncorrelated in the limit $r \to 1$. Hence, when $r$ is close to $1$, one expects to recover the record statistics for the IID model. Thus this RW with resetting model, with parameter $r$, 
nicely interpolates between the two well studied models, namely the IID [$(1-r)\ll 1$ limit] and the RW ($r =0$) cases discussed before (see Fig. \ref{Fig_RG_flow}). We recall that in 
the IID case [$(1-r) \ll 1$ limit], the mean number of records grows universally as $\langle R_N\rangle \sim \ln N$ for large $N$. In the opposite limit $r = 0$ (RW model), the mean number
of records again grows universally, but much faster, as $\langle R_N\rangle \sim \sqrt{4N/\pi}$ for large $N$. It is then interesting to know how $\langle R_N \rangle$ behaves 
for intermediate values of the resetting probability $0\leq r < 1$. Does it stay universal, i.e., independent of the jump distribution $f(\eta)$ as in the RW case?

Here, we compute exactly $\langle R_N \rangle$ for all $0\leq r < 1$ and show that it depends continuously on $r$, nicely interpolating between the two 
known limits. However, to our complete surprise, we found that for every $r$, the average $\langle R_N \rangle$ is completely universal, i.e., independent of the jump distribution 
$f(\eta)$ as long as $f(\eta)$ is symmetric and continuous. Let us briefly summarize our main results. We show that the generating function (GF) of $\langle R_N\rangle$ is given by the exact formula
\bea \label{exact_GF}
\tilde R(z) = \sum_{N=0}^\infty \langle R_N \rangle\, z^N = \frac{\sqrt{1-(1-r)\,z}}{r\,z(1-z)} \ln \left[\frac{1-(1-r)z-r\,z \sqrt{1-(1-r)z}}{1-z} \right] \;,
\eea
which is completely independent of the jump distribution $f(\eta)$. Thus this result holds not only for RW's whose jump distributions have a finite variance, but also for
L\'evy flights where the jump distribution has a fat tail. By analysing this formula (\ref{exact_GF}), we show that, asymptotically for large $N$ and fixed $0< r <1$, the mean
number of records grows as 
\bea \label{RN_largeN}
\langle R_N \rangle = \frac{1}{\sqrt{r}} \, \ln N + \frac{1}{\sqrt{r}} \left( \gamma_E+ \ln(r(1-\sqrt{r}))\right) + O(1/N) \;,
\eea
where $\gamma_E = 0.57721\ldots$ is the Euler constant. Note that this asymptotic large $N$ formula holds strictly for $r<1$. We then consider the two limiting cases $r \to 0$ and $r \to 1$ to see how $\langle R_N \rangle$ converges
respectively to the RW model and the IID model. We show that in the limit $r \to 0$, $N \to \infty$ keeping the product $r\,N$ fixed, $\langle R_N \rangle$ takes the scaling form
\bea \label{f1}
\langle R_N \rangle \approx \frac{1}{\sqrt{r}}\, f_1(r \, N) \;,
\eea
where $f_1(u)$ is a universal scaling function that we compute explicitly [see Eqs. (\ref{f1_explicit}) and~(\ref{f1prime})]. It has the leading asymptotic behaviours
\bea \label{f1_asympt}
f_1(u) \approx
\begin{cases}
&\dfrac{2}{\sqrt{\pi}} \sqrt{u} \;,\;\quad u \to 0 \;,\\
& \ln{u} + \gamma_E\;, \; \, u \to \infty \;.
\end{cases}
\eea
Interestingly, the same scaling function $f_1(u)$ also appears in the expected maximum up to time $t$ of a continuous time Brownian motion with resetting \cite{MMSS21}.
In Section \ref{sec:asympt}, we will make a connection between the mean number of records and the expected maximum. From Eq. (\ref{f1}) and the small $u$ behaviour of $f_1(u)$ in Eq. (\ref{f1_asympt}), one recovers the RW result in Eq. (\ref{av_RN_rw_as}). Similarly, in the opposite limit 
$r\to 1$, $N \to \infty$ keeping the product $(1-r)\,N$ fixed, we show that 
$\langle R_N \rangle$ takes the scaling form
\bea \label{f2}
\langle R_N \rangle \approx f_2((1-r) \, N) \;,
\eea
where $f_2(v)$ is also a universal scaling function given by
\bea \label{f2_expl}
f_2(v) = - {\rm Ei}\left(-\frac{v}{2}\right) + \gamma_E  + \ln\left(\frac{v}{2} \right) \;,
\eea 
where $-{\rm Ei}(-z)=\int_{z}^\infty e^{-t}/t \,dt$. The scaling function has the leading asymptotic behaviours 
\bea \label{f2_asympt}
f_2(v) \approx
\begin{cases}
&\dfrac{v}{2} \;,\;\quad \quad \;\,\,\, \quad \quad \quad \;\;\,v \to 0 \;,\\
& \ln \left(\dfrac{v}{2}\right) + \gamma_E  \;, \;\quad \quad v \to \infty \;.
\end{cases}
\eea
From Eq. (\ref{f2}) and the asymptotic behaviour of $f_2(v)$ as $v \to \infty$ in Eq. (\ref{f2_asympt}), one recovers the leading behaviour $\langle R_N \rangle \approx \ln N$
in the IID model in Eq. (\ref{av_RN_IID_as}). Our numerical simulations are in perfect agreement with our analytical predictions.

The rest of the paper is organized as follows. In Section \ref{sec:gf}, we present the main derivation of the result for the GF of $\langle R_N\rangle$ in Eq. (\ref{exact_GF}) and 
demonstrate precisely how the universality emerges. In Section \ref{sec:asympt}, we derive the asymptotic behaviour of $\langle R_N\rangle$ for large $N$, 
starting from Eq. (\ref{exact_GF}). We also compute the two scaling functions $f_1(u)$ and $f_2(v)$ in the two limits $r\to 0$ and $r \to 1$ respectively. In that Section, we also 
compare our analytical predictions with numerical simulations. Finally, we conclude in Section \ref{sec:conclusion} with some remarks and open questions.

\section{Computing the exact generating function of $\langle R_N \rangle$}\label{sec:gf}

We start with a discrete-time series with $N$ entries $\{x_0=0, x_1, x_2, \cdots, x_N\}$ where the positions $x_n$'s evolve via the stochastic rule in Eq. (\ref{rw_w_reset}), i.e., $x_n$ corresponds to the position of a random walker at step $n$, with a resetting probability $r$ to the origin 
at each step. We recall that a record occurs at step $n$ if $x_n > \max\{x_0=0, x_1, \cdots, x_{n-1}\}$, i.e., it exceeds all the previous positions. We use the
convention that $x_0=0$ is not a record. The reason for this convention will be clear later. We want to compute the average number of records $\langle R_N \rangle$ up to step $N$. To compute this quantity, it is convenient to introduce a binary variable $\sigma_n = \{1,0\}$ at each step such that 
\bea \label{def_sigma}
\sigma_n = 
\begin{cases}
& 1 \quad {\rm if \; a \; record \; occurs \; at \; step \;} n \;,\\
& 0 \quad {\rm otherwise}  \;.
\end{cases}
\eea
Note that, by our convention, $\sigma_0 = 0$. For all $n \geq 1$, the indicator function $\sigma_n$ is a binary random variable that fluctuates from one
realisation of the time series to another. For each realisation of the time series, the number of records $R_N$ (which is also a random variable) can be expressed as a sum 
\bea \label{rel_R_sigma}
R_N = \sum_{n=1}^N \sigma_n \;.
\eea
Taking the average over all realisations, we get
\bea \label{rel_R_sigma_av}
\langle R_N \rangle = \sum_{n=1}^N \langle \sigma_n \rangle \;.
\eea
Since $\sigma_n = \{1,0\}$ is a binary random variable, its average $\langle \sigma_n \rangle$ 
is just the probability that a record occurs at step $n$. Below, we show how to compute $\langle \sigma_n \rangle$ and from it, the mean number of records $\langle R_N \rangle$ using Eq. (\ref{rel_R_sigma_av}).

\begin{figure}
\includegraphics[width = 0.8\linewidth]{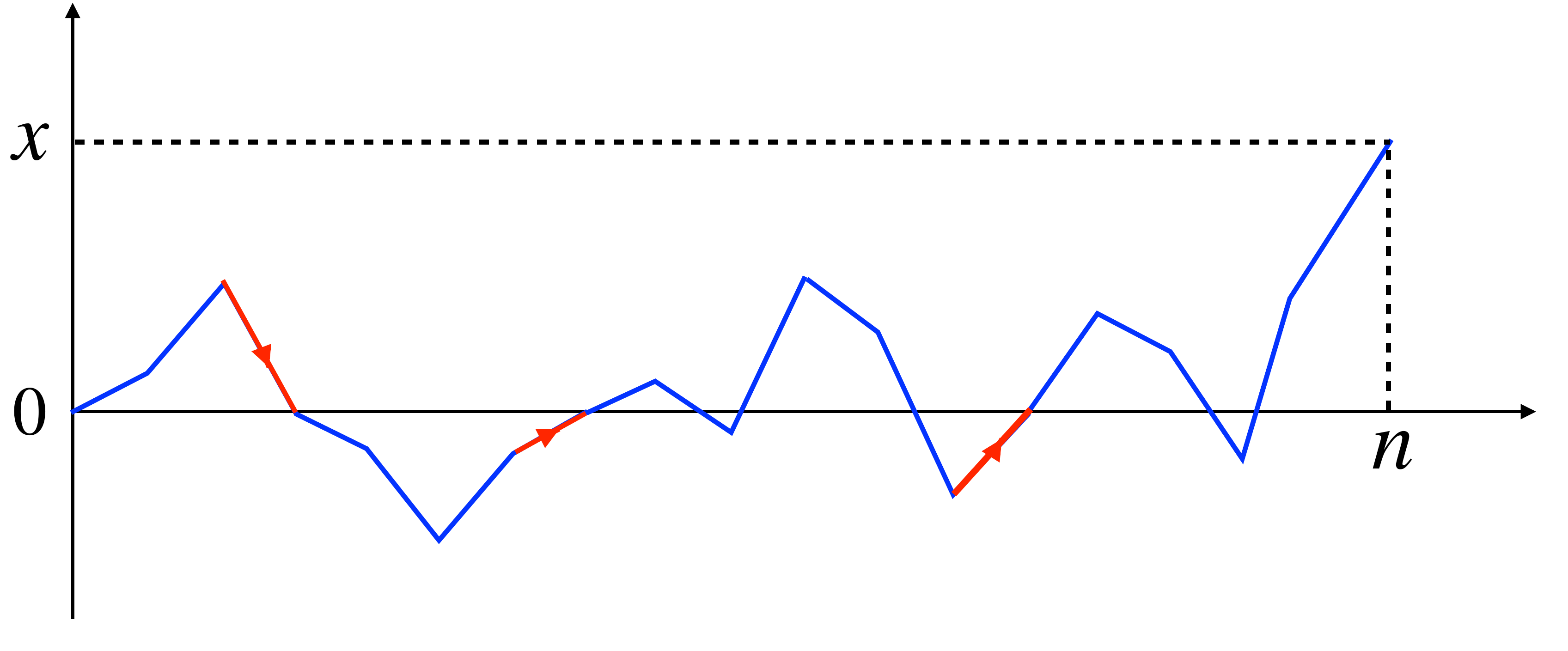}
\caption{Schematic trajectory of a random walk with resetting, starting from $x=0$ up to $n$ steps [see Eq. (\ref{rw_w_reset})]. The resetting events are marked by red arrows. 
If a record happens at step $n$ with value $x$, the trajectory reaches the level $x$ for the first time at step $n$.}\label{Fig_rw}
\end{figure}
Suppose that a record occurs at step $n$, with a record value $x>0$ (see Fig. \ref{Fig_rw}). Let $P_r(x,n)$ denote the probability that the process reaches the level $x$ for the first time at step $n$, starting from $x=0$. Here, the subscript $r$ refers to the presence of a nonzero resetting probability $r$. Hence $\langle \sigma_n \rangle$, the probability that a record occurs at step $n$ can be obtained by integrating $P_r(x,n)$ over all possible $x>0$ at step $n$, i.e.,
\bea \label{ref_sigma_Pr}
\langle \sigma_n \rangle = \lim_{\epsilon \to 0} \int_\epsilon^\infty P_r(x,n)\, dx \;.
\eea 
The reason for putting a cutoff $\epsilon$ is the following. For a record to happen at step $n$, the value of the process at step $n$ must be strictly bigger than $0$, since the process starts at $x=0$. This is true for any process. In the absence of resetting, one can set $\epsilon = 0$ in the outset, because the event that the process reaches exactly $x=0$ at step $n$ is a set with measure zero. However, in the presence of a finite resetting probability $r$, the process resets to $0$ with a nonzero probability and these events should not be counted as records. Hence, we need to impose a finite lower cutoff $\epsilon$ in the integral over $x$ in the integral in Eq. (\ref{ref_sigma_Pr}) (to exclude these ``spurious'' records with value $0$) and eventually take the limit $\epsilon \to 0$. 

Our next step is to estimate $P_r(x,n)$ using a renewal equation approach. For this, we need to first introduce a quantity $Q_0(x,n)$ that denotes the probability that the random walker (without resetting, hence the subscript '$0$'), starting at the origin, stays below the level $x>0$ up to step $n$. We will refer to this as the survival probability of the walker without resetting. Then the renewal equation for $P_r(x,n)$ can be written as
\bea \label{renewal_Pr}
P_r(x,n) = (1-r)^n\, P_0(x,n) + r \, \sum_{m=0}^{n-1} (1-r)^{m} Q_0(x,m) \, P_r(x,n-m-1) \;,
\eea  
starting from $P_r(x,0) = 0$, for all $x>0$. This equation can be understood as follows. Suppose that there is no resetting up to $n$ steps. The probability for this event is $(1-r)^n$. Hence, the first term on the right hand side (rhs) of Eq. (\ref{renewal_Pr}) corresponds to the event of  ``no resetting". The second term on the rhs in Eq. (\ref{renewal_Pr}) corresponds
to the events where the number of resettings within $n$ steps is at least one. In this case, when there is at least one resetting event, suppose that the first resetting event occurs at step $m+1$. The probability for this resetting event is $r(1-r)^m$ since the first $m$ steps have no resetting [this happens with probability $(1-r)^m$] followed by a resetting event exactly 
at step $m+1$ (this occurs with probability $r$). During these first $m$ steps, the random walker evolves without resetting and the probability that this walk stays below the 
level $x$ up to step $m$ is just $Q_0(x,m)$. Finally, after the first resetting at step $m+1$, the process renews itself, starting from the origin, and eventually arrives at $x$ for the first time at step $n$ of the original process, i.e., at step $n-m-1$ for the renewed process starting at $0$ at step $m+1$. Thus the entire probability of the trajectory with first resetting at step $m+1$ is given by the product $r\, (1-r)^{m} Q_0(x,m) \, P_r(x,n-m-1)$. Finally, noting that the first resetting can occur at all intermediate steps between $0$ and $n$ we obtain the result in Eq. (\ref{renewal_Pr}) by summing over all possible values of $m$. Notice that Eq. (\ref{renewal_Pr}) holds for all $x \geq 0$ and all $n\geq 0$. For $n=0$, we just interpret the sum in the
second term in Eq.~(\ref{renewal_Pr}) as $0$.

We next define the generating functions 
\bea \label{def_GFs}
\tilde P_r(x,z) = \sum_{n=0}^\infty P_r(x,n)\, z^n \quad \quad {\rm and} \quad \quad \tilde Q_0(x,z) = \sum_{m=0}^\infty Q_0(x,m)\, z^m  \;.
\eea
Multiplying both sides of Eq. (\ref{renewal_Pr}) by $z^n$ and summing over $n$ from $n=0$ to $n \to \infty$, we obtain, upon using the convolution structure of the second term in Eq. (\ref{renewal_Pr})
\bea \label{renewal_GF1}
\tilde P_r(x,z) = \tilde P_0(x,(1-r)z) + r\,z\, \tilde Q_0(x,(1-r)z) \, \tilde P_r(x,z) \;.
\eea
This allows us to express the GF $\tilde P_r(x,z)$ in terms of the generating functions without resetting ($r=0$), namely
\bea \label{renewal_GF2}
\tilde P_r(x,z) = \frac{\tilde P_0(x,(1-r)\,z)}{1-r\,z\, \tilde Q_0(x,(1-r)z)} \;.
\eea 
For a standard random walk/L\'evy flight without resetting, the first-passage probability $P_0(x,n)$ to level $x$ at step $n$ and the survival probability $Q_0(x,n)$ 
to stay below the level $x$ up to step $n$ are well studied \cite{Poll52,SA1954,Spi56,Iva94,CM2005,Satya2010,MMS2017}. We first note that $Q_0(x,n)$ coincides with the probability that the random walker, starting at $x>0$ initially, does not cross $0$ up step $n$. This is easily seen by making a change of coordinates $y_n = x-x_n$. Similarly, $P_0(x,n)$ can be identified with the probability that the random walker, starting at the origin, reaches the level $x>0$ at step $n$, without crossing the origin in between. This is seen first by making the transformation $y_n = x - x_n$,
followed by the reversal of the time. The GF's of these two probabilities for a random walk can, in principle, be extracted from their Laplace transforms, which are known to satisfy \cite{Iva94,CM2005,WMS2012,MMS2014,MMS2017}
\bea 
&&\int_0^\infty \tilde P_0(x,z)\, e^{-\lambda \,x}\, dx = \phi(\lambda,z) \label{LT_GF1} \\
&&\int_0^\infty \tilde Q_0(x,z)\, e^{-\lambda \,x}\, dx = \frac{\phi(\lambda,z)}{\lambda \sqrt{1-z}} \label{LT_GF2} 
\eea 
where
\bea \label{def_phi}
\phi(\lambda,z) = \exp{\left(-\frac{\lambda}{\pi} \int_{-\infty}^\infty \frac{\ln(1-z \hat f(k))}{\lambda^2 + k^2} dk \right)} \;,
\eea
where $\hat f(k) = \int_{-\infty}^\infty f(x)\,e^{i k x}\, dx$ is the Fourier transform of the jump distribution. Note that the dependence on the jump distribution $f(\eta)$
occurs only through the function $\phi(\lambda,z)$. Interestingly, the dependence on the jump distribution $f(\eta)$ of $\tilde Q_0(x,z)$ in Eq. (\ref{LT_GF2}) completely
disappears in the limit $x \to 0$. To see this, one can perform the change of variable $\lambda x = u$ in Eq. (\ref{LT_GF2}) and take  the limit $\lambda \to \infty$, using
$\phi(\lambda \to \infty, z) = 1$. This gives
\bea \label{SA}
\tilde Q_0(0,z) = \frac{1}{\sqrt{1-z}} = \sum_{n=0}^\infty {2n \choose n}2^{-2n}\, z^n\;.
\eea  
Thus $Q_0(0,n) = {2n \choose n}2^{-2n}$ is universal for any $n$, i.e., independent of the jump distribution $f(\eta)$, as long as it is symmetric and continuous. This is known as the celebrated Sparre Andersen theorem \cite{SA1954}. Similarly, in the limit $x \to \infty$ one can extract $Q_0(x \to \infty, z)$ by taking the $\lambda \to 0$ limit in Eq. (\ref{LT_GF2}). Note that in the limit $\lambda \to 0$, one obtains $\phi(\lambda \to 0,z) = 1/\sqrt{1-z}$ \cite{SA1954}. This gives
\bea \label{Qxinf}
\tilde Q_0(x \to \infty,z) = \frac{1}{1-z} = \sum_{n=0}^\infty z^n\;.
\eea
Hence $Q_0(x \to \infty,n) = 1$ for all $n \geq 0$, which of course is expected since, if the walker starts at $\infty$   
it does not cross $0$ up to step $n$ with probability $1$. Thus, in both limits $x\to 0$ and $x \to \infty$, the GF of the survival probability $\tilde Q_0(x,z)$ is universal. 
We will see shortly that these two limiting universal behaviours will play an important role in establishing the universality of the mean number of records.

\begin{figure}[ht]
\includegraphics[width=0.5\linewidth]{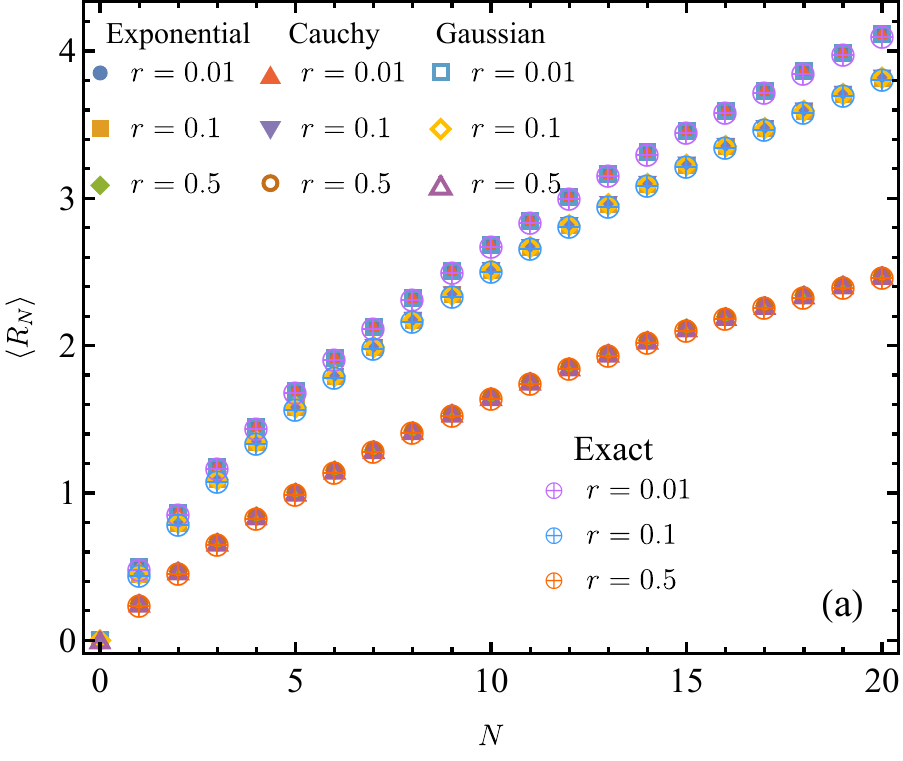}\includegraphics[width=0.52\linewidth]{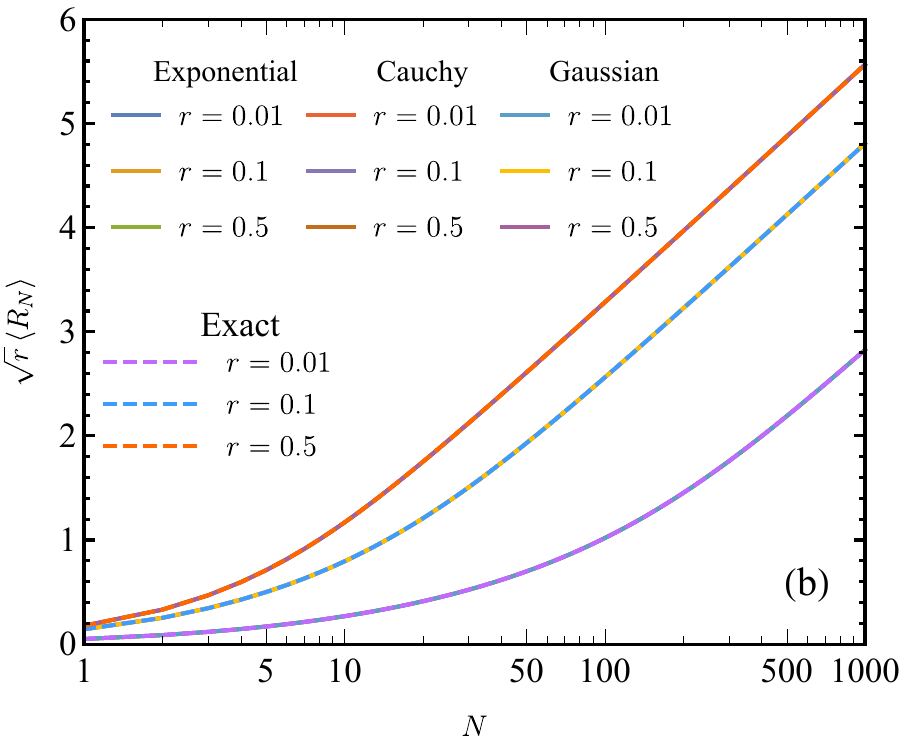}
\caption{Plot of the average number of records $\langle R_N \rangle$ vs $N$ for three different values of $r=0.01, 0.1$ and $r=0.5$. For each value of $r$, the numerical data for three different jump distributions coincide for all $N$. In panel a), we also show the exact results for $\langle R_N \rangle$ for smaller values of $N$ extracted from Eq.~(\ref{GF_R1}) by expanding the GF in powers of $z$ using Mathematica. The first three values ($N=1, 2$ and $3$) are given in Eq. (\ref{RN_expl}). We find a perfect agreement between the numerical results and the analytical predictions. This also confirms the universality for even smaller values of $N$. In panel b), we plot the analytical and the numerical results for $\langle R_N \rangle$ vs $N$ for a wider range of $N$ (plotted on a log-linear scale) and for three different values of $r$. For each $r$, we see that the three different jump distributions give identical results for all $N$, thus confirming the universality.}\label{Fig_exact}
\end{figure}

There exists a nice relation between $\tilde P_0(x,z)$ and $\tilde Q_0(x,z)$ which will be useful shortly. To derive this relation, we integrate Eq. (\ref{LT_GF2}) by parts and we use the Sparre Andersen theorem~(\ref{SA}), namely $\tilde Q_0(0,z) = {1}/{\sqrt{1-z}}$. This gives
\bea\label{rel_PQ}
1 + \sqrt{1-z} \int_0^\infty \frac{\partial \tilde Q_0(x,z)}{\partial x}\, e^{-\lambda x}\, dx = \phi(\lambda, z) = \int_0^\infty \tilde P_0(x,z)\, e^{-\lambda\,x}\, dx \;,
\eea
where in the last equality we used the relation (\ref{LT_GF1}). Comparing the rhs and the lhs of Eq. (\ref{rel_PQ}), we get
\bea \label{rel_PQ2}
\tilde P_0(x,z) = \delta(x) + \sqrt{1-z}   \;\frac{\partial \tilde Q_0(x,z)}{\partial x} \;.
\eea 
Substituting this relation in Eq. (\ref{renewal_GF2}), we obtain
\bea \label{rel_PQ3}
\tilde P_r(x,z) =  \frac{\delta(x) +   \sqrt{1-z}   \; \partial_x \tilde Q_0(x,(1-r)z) }{1-r\,z\, \tilde Q_0(x,(1-r)z)} \;.
\eea
We now substitute this expression for $\tilde P_r(x,z)$ on the rhs of Eq. (\ref{ref_sigma_Pr}). Due to a finite lower cut-off $\epsilon$, the term containing $\delta(x)$ in Eq. (\ref{rel_PQ3}) does not contribute to the integral. Moreover, the second term can be integrated explicitly yielding
\bea \label{GF_sigma}
\sum_{n=0}^\infty \langle \sigma_n \rangle\, z^n = \frac{\sqrt{1-(1-r)z}}{r\,z} \;  \ln \left( \frac{1-r\,z\, \tilde Q_0(0,(1-r)z)}{1-r\,z\, \tilde Q_0(x\to\infty, (1-r)z)}\right) \;.
\eea
Interestingly, the rhs of Eq. (\ref{GF_sigma}) involves only the limiting values of the GF $\tilde Q_0(x,(1-r)z)$, namely the $x \to 0$ and $x \to \infty$ limits. As shown before, $\tilde Q_0(x,(1-r)z)$ is universal in both these limits. Consequently, the rhs of Eq. (\ref{GF_sigma}) is universal, i.e., independent of $f(\eta)$ as long as it is symmetric and continuous.    
Indeed, using the results in Eqs. (\ref{SA}) and (\ref{Qxinf}), we get
\bea \label{GF_sigma2}
\sum_{n=0}^\infty \langle \sigma_n \rangle\, z^n =
\frac{\sqrt{1-(1-r)\,z}}{r\,z} \ln \left[\frac{1-(1-r)z-r\,z \sqrt{1-(1-r)z}}{1-z} \right] \;.
\eea
Finally, using the relation (\ref{rel_R_sigma_av}), one gets the GF of the mean number of records $\langle R_N \rangle$
\begin{equation} \label{GF_R1}
\sum_{N=0}^\infty \langle R_N \rangle\, z^N = \frac{1}{1-z} \sum_{n=0}^\infty \langle \sigma_n \rangle\, z^n =  \frac{\sqrt{1-(1-r)\,z}}{r\,z(1-z)} \ln \left[\frac{1-(1-r)z-r\,z \sqrt{1-(1-r)z}}{1-z} \right] \;.
\end{equation}
Thus, remarkably, the dependence on $\phi(\lambda,z)$, and hence on $f(\eta)$, completely drops out and we get a universal result for any $0\leq r< 1$. Note that $\langle R_N \rangle$ is universal, i.e., independent of $f(\eta)$ for all $N$, and not just for large $N$. For example, by expanding the rhs of (\ref{GF_R1}) in powers of $z$, one can easily obtain $\langle R_N \rangle$ for the first few values of $N$ such as $N=1, 2$ and~$3$  
\bea \label{RN_expl}
\langle R_1 \rangle = \frac{1-r}{2} \quad, \quad \langle R_2 \rangle =\frac{1}{8} \left(-r^2-6 r+7\right) \quad, \quad \langle R_3 \rangle= \frac{1}{16} \left(-r^3-3 r^2-15 r+19\right) \;.
\eea
For general $N$, it is hard to extract $\langle R_N \rangle$ explicitly from its GF in Eq. (\ref{GF_R1}). However, as shown in the next Section, 
one can easily extract $\langle R_N \rangle$ for large $N$, by analysing the limit $z \to 1$ of the GF in Eq. (\ref{GF_R1}).

In Fig. \ref{Fig_exact}, we test the universality of $\langle R_N \rangle$ vs $N$ for all $N$ by computing $\langle R_N\rangle$ numerically for three different values of $r$ and, for each $r$, for three different jump distributions: (i) Exponential, $f(\eta) = (1/2) e^{-|\eta|}$ (for simplicity we call it ``exponential'', even though it is actually a double-sided exponential distribution), (ii) Cauchy, $f(\eta) = (1/\pi)1/(1+\eta^2)$ and (iii) Gaussian, $f(\eta) = (1/\sqrt{2\pi})e^{-x^2/2}$. For each $r$, we see that the results for the three distributions coincide with each other and also with the exact analytical answer extracted from the GF in Eq. (\ref{GF_R1}).

\section{Asymptotic expansion for large $N$}\label{sec:asympt}

In this section, we first analyse $\langle R_N \rangle$ for large $N$ with $r$ fixed. Then, we investigate the scaling limit $N \to \infty$ and $r \to 0$ with the product $u=r\, N$ fixed, to describe the crossover from the RW limit when $r$ increases from $0$. In this limit, we also show how the result for $\langle R_N \rangle$ can be translated to the expected maximum of a continuous time Brownian motion with resetting. Finally, we also investigate the opposite scaling limit $N \to \infty$ and $r \to 1$ with the product $u=(1-r)\,N$ fixed, to describe the crossover from the IID limit, when $r$ decreases from $1$.

\subsection{The limit $N \to \infty$ for fixed $r$}

\begin{figure}
\centering
\includegraphics[width = 0.5 \linewidth]{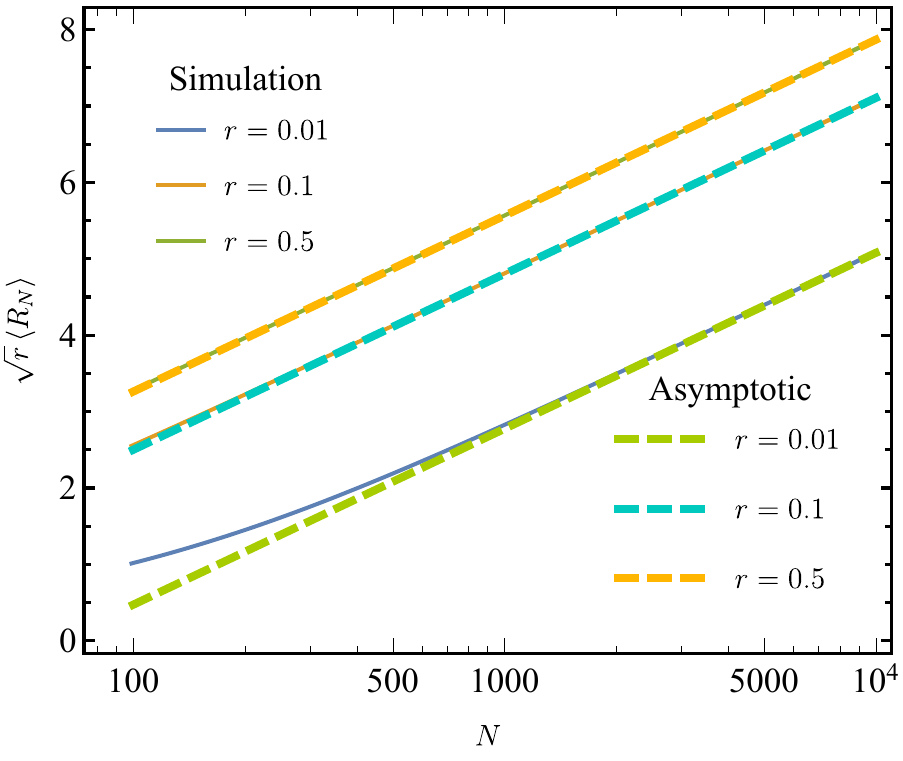}
\caption{The numerically obtained $\langle R_N \rangle$ vs $N$ for large $N$ for the exponential jump distribution $f(\eta) = (1/2) e^{-|\eta|}$ for $r=0.01, r=0.1$ and $r=0.5$ compared with the asymptotic formula in Eq.~(\ref{RNlargeN}), where we kept only the two leading order terms for large $N$. The asymptotic result in Eq.~(\ref{RNlargeN}) is expected to hold
only for very large $N$, i.e., $N > N^*(r)$ where $N^*(r)$ is a crossover value. From the data for $r=0.01$ we see that $N^*(r=0.01) \approx 1000$, below which one needs to keep higher order terms in Eq. (\ref{RNlargeN}).}\label{Fig_asym}
\end{figure}
Here, we set $z = 1 - s$ with $s \to 0$ in Eq. (\ref{GF_R1}). The left hand side (lhs) then converges to the Laplace transform 
\bea \label{conv_LT}
\sum_{N=0}^\infty \langle R_N\rangle \, z^N \approx \int_0^\infty  \langle R_N\rangle\,e^{-s N}\, dN \;.
\eea
On the other hand, the rhs of Eq. (\ref{GF_R1}), to leading orders in $s \to 0$ behaves as
\bea \label{rhs_small_s}
{\rm rhs}  \approx - \frac{1}{\sqrt{r}}\, \frac{\ln s}{s} + \left[\frac{1}{\sqrt{r}} \ln{(r-r^{3/2})}\right] \frac{1}{s} \;.
\eea
Equating the lhs and the rhs gives
\bea \label{comp}
 \int_0^\infty  \langle R_N\rangle\,e^{-s N}\, dN \approx  - \frac{1}{\sqrt{r}}\, \frac{\ln s}{s} + \left[\frac{1}{\sqrt{r}} \ln{(r-r^{3/2})}\right] \frac{1}{s} \;.
\eea
Using
\bea \label{inv_LT}
{\cal L}^{-1}_{s\to N} \left(-\frac{\ln s}{s}\right) = \ln N + \gamma_E \;, 
\eea
one can easily invert the Laplace transform in Eq. (\ref{comp}) to get the leading asymptotic behaviour of $\langle R_N \rangle$ for large $N$ 
\bea \label{RNlargeN}
\langle R_N \rangle = \frac{1}{\sqrt{r}} \, \ln N + \frac{1}{\sqrt{r}} \left( \gamma_E+ \ln(r(1-\sqrt{r}))\right) + O(1/N) \;.
\eea
In Fig. \ref{Fig_asym}, we compare this asymptotic result to the numerically obtained $\langle R_N \rangle$ for three different values of $r$ and for the exponential 
jump distribution $f(\eta) = (1/2) e^{-|\eta|}$. We have also checked (not shown in the figure to avoid crowding) that, for a fixed $r$, exactly the same results are obtained 
for each $N$, for a Cauchy and a Gaussian jump distribution, as expected from the universality.

\subsection{The limit $N \to \infty$, $r \to 0$ keeping $r\,N$ fixed}

To study the large $N$ limit, we set, as before, $z=1-s$ with $s \to 0$. The lhs of Eq. (\ref{GF_R1}) converges, as
before to the Laplace transform in Eq. (\ref{conv_LT}). We anticipate a scaling form as in Eq. (\ref{f1}) -- verified a posteriori -- 
\bea \label{scaling_f1_txt}
\langle R_N \rangle \approx \frac{1}{\sqrt{r}}\, f_1(r \, N) \;.
\eea
Substituting this scaling form in the Laplace transform in Eq. (\ref{conv_LT})
and making a change of variable $r\, N = u$, we get 
\bea \label{conv_LTf1} 
\sum_{N=0}^\infty \langle R_N\rangle \, z^N \approx \int_0^\infty  \langle R_N\rangle\,e^{-s N}\, dN \approx \frac{1}{r^{3/2}} \int_0^\infty f_1(u)\,e^{-\frac{s}{r}u}\, du \;.
\eea
Since we are considering the $r \to 0$ limit, and $s \to 0$ limit simultaneously, we see that we need to keep the ratio $s/r = p$ fixed in Eq. (\ref{conv_LTf1}). This gives
\bea \label{conv_LTf1_2} 
\sum_{N=0}^\infty \langle R_N\rangle \, z^N \approx \frac{1}{s^{3/2}} \left[p^{3/2} \int_0^\infty f_1(u)\, e^{-pu}\, du \right] \;.
\eea
On the rhs of Eq. (\ref{GF_R1}), we set $r=s/p$ with $p$ fixed to be of $O(1)$ and expand for small $s$. This gives the small $s$ behaviour of the rhs of Eq. (\ref{GF_R1}) for fixed $p$ as
\bea \label{rhs_f1}
{\rm rhs} \approx \frac{1}{s^{3/2}}\left[ \sqrt{p(1+p)}\ln{\left( \frac{1+p}{p}\right)}\right]  \;.
\eea
Equating Eqs. (\ref{conv_LTf1_2}) and (\ref{rhs_f1}) gives us the exact Laplace transform of the scaling function $f_1(u)$ 
\bea \label{exact_LT_f1} 
\int_0^\infty f_1(u)\, e^{-pu}\, du = \frac{\sqrt{1+p}}{p} \ln{\left( \frac{1+p}{p}\right)} \;.
\eea
\begin{figure}[t]
\includegraphics[width=0.5\linewidth]{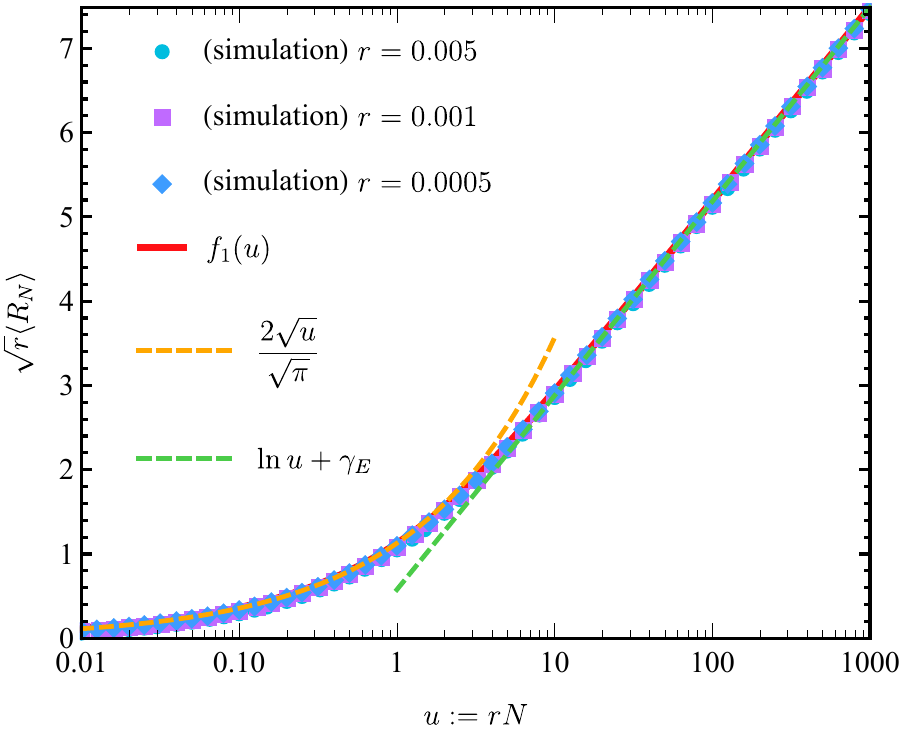}
\caption{Numerical verification of the scaling form in Eq. (\ref{scaling_f1_txt}): $\sqrt{r} \, \langle R_N \rangle$ plotted vs $u=r\,N$ for three different small values of $r$ and different values of large $N$. All the three curves collapse onto the same master curve, demonstrating the validity of the scaling form in Eq. (\ref{scaling_f1_txt}). Furthermore, the numerically obtained scaling function matches perfectly with the analytical scaling function $f_1(u)$ in Eq. (\ref{f1_explicit}). The orange and the green dashed curves indicate the asymptotic behaviours for $u \to 0$ and $u \to \infty$ respectively in Eq. (\ref{asympt_f1}). }\label{Fig_f1}
\end{figure}
To invert this Laplace transform, we can use the following two identities 
\bea 
&&{\cal L}^{-1}_{p\to t} \left(\frac{\sqrt{p+1}}{p} \right) = \frac{1}{\sqrt{\pi t}}\,e^{-t} + {\rm erf}(\sqrt{t}) \label{LT_inv_f11} \\
&&{\cal L}^{-1}_{p\to t}  \ln{\left( \frac{1+p}{p}\right)} = \frac{1}{t}(1-e^{-t})  \label{LT_inv_f12} \;,
\eea
where ${\rm erf}(z) = (2/\sqrt{\pi}) \int_{0}^z e^{-u^2}\,du$ is the error function. Now using the convolution theorem, we can explicitly write the scaling function (which is universal) as
\bea \label{f1_explicit}
f_1(u) = \int_0^u dt\, \left(\frac{1}{\sqrt{\pi t}}\,e^{-t} + {\rm erf}(\sqrt{t})  \right)\left( \frac{1-e^{-(u-t)}}{u-t}\right) \;.
\eea
This integral seems hard to perform explicitly. However, its derivative $f_1'(u)$ can be expressed in terms of a hypergeometric function as shown in the Appendix \ref{App},
\bea \label{f1prime}
f_1'(u) = \frac{e^{-u}}{\sqrt{\pi\,u}} \,_2 F_2\left( \{1,1\}\,, \{\frac{1}{2}, 2\},u\right) \;.
\eea
The asymptotic behaviour of $f_1'(u)$ can be easily obtained from the hypergeometric function and one gets $f_1'(u) \approx 1/\sqrt{u}$ as $u \to 0$ and $f_1'(u) \approx 1/u$ as $u \to \infty$. Integrating over $u$, and using $f_1(0) = 0$, one gets the following asymptotic behaviour of the scaling function 
\bea \label{asympt_f1}
f_1(u) \approx
\begin{cases}
&\dfrac{2}{\sqrt{\pi}}\,\sqrt{u} \quad, \quad \quad \quad u \to 0\;, \\
& \ln u + \gamma_E \quad, \quad \quad \; u \to \infty \;.
\end{cases}
\eea
Note that the constant correction term $\gamma_E$ for large $u$ is not easy to obtain from the large $u$ expansion of $f_1'(u)$. However, it can be obtained by analysing
the small $p$ behaviour of the Laplace transform in Eq. (\ref{exact_LT_f1}), where it behaves as $-(\ln p)/p$ for small $p$. In Fig. \ref{Fig_f1}, we compare the analytical scaling function $f_1(u)$ with the numerically obtained scaling function, finding an excellent agreement.

If one now substitutes the small $u$ asymptotics in the scaling form in Eq. (\ref{scaling_f1_txt}), one gets, as $r \to 0$ (for large $N$), $\langle R_N\rangle \approx (2/\sqrt{\pi})\sqrt{N}$, thus
recovering the leading asymptotic result for random walk without resetting in Eq. (\ref{av_RN_rw_as}). In contrast, in the limit $u \to \infty$ (i.e. $r$ small but fixed and $N \to \infty$), substituting the large $u$ asymptotic behaviour of $f_1(u)$ in the scaling form in Eq. (\ref{scaling_f1_txt}) gives
\bea \label{f1_largeu}
\langle R_N \rangle \approx \frac{1}{\sqrt{r}}\, \ln(r\,N)  \approx \frac{1}{\sqrt{r}}\, \ln N\;,
\eea
which matches smoothly with the leading order for large $N$ (for fixed $r$) in Eq. (\ref{RNlargeN}).

Note that this $\ln N$ asymptotics of $\langle R_N \rangle$ for large $N$ in Eq. (\ref{f1_largeu}) is reminiscent of the IID model in Eq. (\ref{av_RN_IID_as}), however with an important difference that
the prefactor $1/\sqrt{r}$ still carries information about the correlations between the entries in the $r\to 0$ limit. In fact, this behaviour $\langle R_N \rangle \approx ({1}/{\sqrt{r}})\, \ln(r\,N)$ can be understood from a physical ``block argument'' as follows. We consider the positions of the 
random walker with resetting in the $r \to 0$ limit. In this limit, the time interval $n$ between two successive resetting events is exponentially distributed as $r\, (1-r)^n \sim r\, e^{-n\,r}$, with correlation time $\tau=1/r$, which is very large. We now use a ``real space'' renormalisation group argument discussed before in the context of extreme value statistics \cite{MPS2020}. This consists of the following steps. It is convenient to compute first $\langle \sigma_N \rangle$, i.e., the probability that a record occurs at the last step $N$. This is the same event that the last entry is the global maximum of the time series. 
\begin{itemize}

\item[(i)] {We first divide the $N$ entries (i.e., the positions of the walker) into blocks of size  of the order of the correlation time $\tau = 1/r$. So, the number of blocks is $N_b \approx N/(1/r) = r\,N$. In general, $N_b$ is a random variable but for large $N$ one can replace it by its average value, which is indeed given by $r \, N$ for small $r$.}

\item[(ii)] The entries from block to block are independent. Hence, the probability that the global maximum will belong to the last block is $1/N_b = 1/(rN)$.

\item[(iii)] Given that the global maximum belongs to the last block: what is the probability $p(r)$ that the global maximum
is actually the last entry (at step $N$ of the full time series) of the last block, so that it is a record? One can easily compute this probability $p(r)$
as follows. In the last block, we have a pure random walk of say $m$ steps, i.e, $(m+1)$ entries. The probability that the last
entry (within the last block of $m+1$ steps) is a record is known to be given by the survival probability $Q_0(x=0,m)$ \cite{WMS2012}, which is given by
the Sparre Andersen theorem $Q_0(x=0,m) = {2m \choose m}2^{-2m}$ [see Eq. (\ref{SA})]. However, the size
$m$ of the last block is a random variable distributed via $r(1-r)^m$, where $m = 0, 1, 2, \ldots$ (i.e., resetting at the first entry
followed by no resetting during $m$ consecutive steps). Hence, the probability that the last entry of the last block is a record
within the block is given by
\bea \label{pr}
p(r) = \sum_{m=0}^\infty r(1-r)^m {2m \choose m}2^{-2m} = \frac{r}{\sqrt{1-(1-r)}} = \sqrt{r} \;,
\eea
where we used the result that $\sum_{m=0}^\infty {2m \choose m}2^{-2m}\,z^m = 1/\sqrt{1-z}$ as in Eq. (\ref{SA}). 

\item[(iv)] Hence, the record rate, i.e. $\langle \sigma_N \rangle$, at step $N$ (which is the probability that a record happens at step $N$) 
is therefore
\bea \label{arg_RG}
\langle \sigma_N \rangle &=& {\rm Prob. \; [the \; global \; maximum \; belongs \; to \; the \; last \; block]} \nonumber \\
&\times& {\rm Prob. \; [the \; global \; maximum \; occurs \; at \; the \; last \; step \; of \; the \; last \; block]} \nonumber \\
&\approx& \frac{1}{N_b} \sqrt{r}  \approx \frac{1}{\sqrt{r}} \frac{1}{N} \;,
\eea
where we used $N_b \approx r\,N$. Finally, using $\langle R_N \rangle - \langle R_{N-1} \rangle = \langle \sigma_N \rangle$, we get the result $\langle R_N \rangle \approx (1/\sqrt{r}) \ln N$, as obtained 
exactly in Eq. (\ref{f1_largeu}). Thus this argument clearly shows that the prefactor $1/\sqrt{r}$ comes from the correlation time $1/r$.

\end{itemize}

\vspace*{0.5cm}

\noindent{\it Connection between the mean number of records and the average maximum of 
a random walk with resetting.} We have seen above that in the limit $r \to 0$, $N \to \infty$, keeping $r\,N$ fixed,
$\langle R_N \rangle \approx (1/\sqrt{r})\, f_1(r\, N)$ where $f_1(u)$ is given in Eq. (\ref{f1_explicit}). It is convenient to rewrite
it as
\bea \label{f1_tilde}
\langle R_N \rangle \approx \sqrt{N} F_1\left(r\,N \right) \quad, \quad {\rm where} \quad F_1(u) = \frac{1}{\sqrt{u}} f_1(u) \;. 
\eea  
For the same random walk with resetting (\ref{rw_w_reset}), another interesting observable is $M_N = \max \{x_0=0, x_1, \cdots, x_N \}$. 
One can ask: what is the average maximum $\langle M_N \rangle$? Intuitively, one may expect that this may be related to the mean number of
records $\langle R_N \rangle$. For a discrete time lattice random walks with jumps $\pm 1$, it is easy to see that $\langle R_N \rangle = \langle M_N \rangle$,
since whenever a new record happens, both $R_N$ and $M_N$ increase by $1$ \cite{WMS2012}. In contrast, for a random walk with continuous 
jump distribution $f(\eta)$ with a finite variance $\sigma^2$, it was shown in Ref.~\cite{WMS2012} that these two observables are proportional, but not
exactly equal, namely
\bea \label{rel_RN_MN}
\langle M_N \rangle \approx \frac{\sigma}{\sqrt{2}}  \langle R_N \rangle \;.
\eea
On general grounds, one would conjecture this relation to be still valid in the presence of resetting. Below we show that this conjecture
is certainly true in the scaling limit $r \to 0$, $N \to \infty$, keeping $r\,N$ fixed. 

Indeed, assuming that this conjecture (\ref{rel_RN_MN}) is true, 
substituting the scaling form (\ref{f1_tilde}) in Eq. (\ref{rel_RN_MN}), one would predict that 
\bea \label{prediction_MN}
\langle M_N \rangle \approx \sqrt{\frac{N \sigma^2}{2}}\, F_1\left( r\, N\right) \;.
\eea
In this scaling limit, the random walk with resetting converges to the continuous time Brownian motion with resetting once we identify
\bea \label{cont_limit}
r = \tilde r \,\Delta t \quad , \quad N = \frac{t}{\Delta t}  \quad {\rm and} \quad \sigma^2 = 2 D \, \Delta t \;,
\eea
where $\tilde r$ is the resetting rate and $t$ denotes the continuous time in units of $\Delta t \to 0$. In addition, one needs to scale the variance as above where
$D$ denotes the diffusion constant of the Brownian motion. Substituting these relations (\ref{cont_limit}) in Eq. (\ref{prediction_MN}) one would get the prediction that, for a continuous time Brownian motion with resetting rate $\tilde r$, the expected maximum $\langle M(t) \rangle$ up to time $t$ would be given by
\bea \label{Moft}
\langle M(t) \rangle \approx \sqrt{D\,t} \, F_1(\tilde r\,t) = \sqrt{\frac{D}{\tilde r}} \, f_1(\tilde r\,t) \;,
\eea 
with $f_1(u)$ given in Eq. (\ref{f1_explicit}). Note that for a continuous time Brownian motion, this relation is expected to hold for all $t$ and
all $\tilde r>0$. In fact, the expected maximum for a continuous time Brownian motion was recently computed in Ref. \cite{MMSS21} in the context of
computing the mean perimeter of the convex hull of a $2d$ resetting Brownian motion and the result in (\ref{Moft}) coincides exactly with that
in Ref. \cite{MMSS21}. This validates, a posteriori, the conjecture (\ref{rel_RN_MN}) for discrete-time random walker, at least in this scaling limit $r\to 0$, $N \to \infty$, 
keeping $r\,N$ fixed, where it converges to continuous time Brownian motion. It will be interesting to prove this conjecture (\ref{rel_RN_MN}) for a discrete-time
random walker (with a finite variance of the continuous and symmetric jump distribution) for arbitrary $0<r<1$.

\subsection{The limit $N \to \infty$, $r \to 1$ keeping $(1-r)\,N$ fixed}

\begin{figure}[t]
\includegraphics[width=0.5\linewidth]{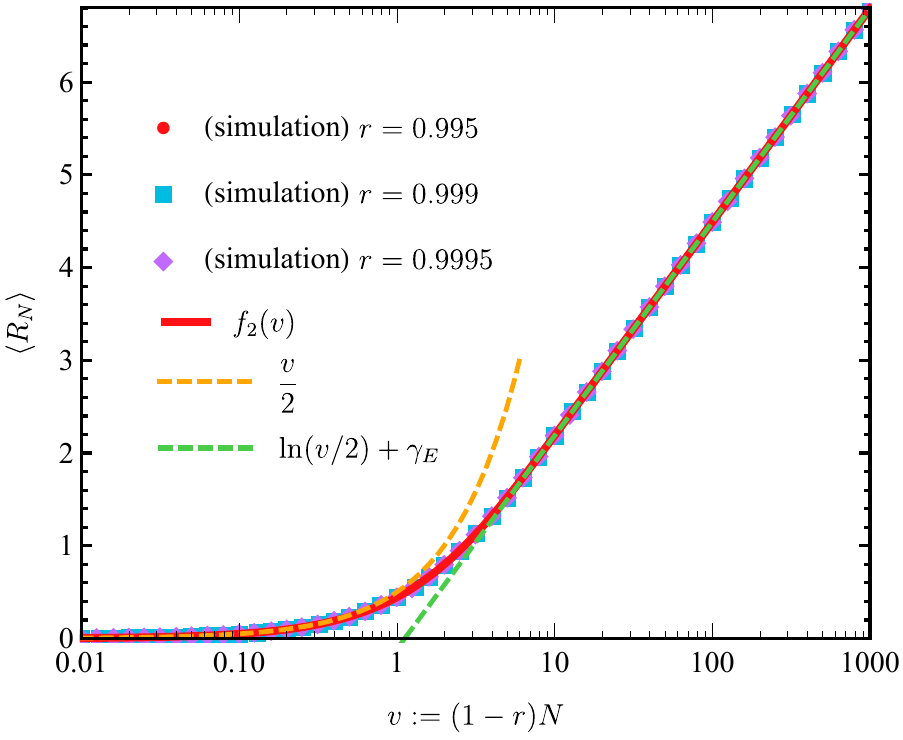}
\caption{Numerical verification of the scaling form in Eq. (\ref{scaling_f2_txt}): $\langle R_N \rangle$ plotted vs $v=(1-r)\,N$ for three different small values of $r$ and different values of large $N$. All the three curves collapse onto the same master curve, demonstrating the validity of the scaling form in Eq. (\ref{scaling_f2_txt}). Furthermore, the numerically obtained scaling function matches perfectly with the analytical scaling function $f_2(v)$ in Eq. (\ref{f2_expl}). The orange and the green dashed curves indicate the asymptotic behaviours $v \to 0$ and $v \to \infty$ respectively in Eq. (\ref{f2_asympt}). }\label{Fig_f2}
\end{figure}
As before, we start from the exact expression for the GF in Eq. (\ref{GF_R1}) and set $z=1-s$ with $s \to 0$. In the $r \to 1$ limit, and $N \to \infty$ limit, we now anticipate
a scaling form, again justified a posteriori, 
\bea\label{scaling_f2_txt}
\langle R_N \rangle \approx f_2((1-r)\,N) \;.
\eea
Substituting this scaling form on the rhs of Eq. (\ref{conv_LT}), we get
\bea \label{conv_LTf2} 
\sum_{N=0}^\infty \langle R_N\rangle \, z^N \approx \int_0^\infty  \langle R_N\rangle\,e^{-s N}\, dN \approx \frac{1}{1-r} \int_0^\infty f_2(v)\,e^{-\frac{s}{1-r}v}\, dv \;.
\eea
Since we are considering the $r \to 1$ limit, and $s \to 0$ limit simultaneously, we see that we need to keep the ratio $s/(1-r) = p$ fixed in Eq. (\ref{conv_LTf1}). This gives
\bea \label{conv_LTf2_2} 
\sum_{N=0}^\infty \langle R_N\rangle \, z^N \approx \frac{1}{s} \left[p \int_0^\infty f_2(v)\, e^{-pv}\, dv \right] \;.
\eea
On the rhs of Eq. (\ref{GF_R1}), we set $(1-r)=s/p$ with $p$ fixed to be of $O(1)$ and expand for small~$s$. This gives the small $s$ behaviour of the rhs of Eq. (\ref{GF_R1}) for fixed $p$ as
\bea \label{rhs_f2}
{\rm rhs} \approx \frac{1}{s}\left[ \ln{\left(\frac{1+2p}{2p}\right)}\right]  \;.
\eea
Equating Eqs. (\ref{conv_LTf2_2}) and (\ref{rhs_f2}) gives us the exact Laplace transform of the scaling function~$f_2(v)$ 
\bea \label{exact_LT_f2} 
\int_0^\infty f_2(v)\, e^{-pv}\, dv = \frac{1}{p}\, \ln \left(\frac{1+2p}{2p}\right) \;.
\eea
This Laplace transform can be explicitly inverted to give the result in Eq. (\ref{f2_expl}), with the asymptotic behaviours given in Eq. (\ref{f2_asympt}).

\section{Summary and Conclusion}\label{sec:conclusion}

In this paper, we have studied the record statistics of a random walk of $N$ steps with a resetting probability $r \in [0,1[$. This model interpolates between 
two well studied models, namely the standard random walk model in the limit $r \to 0$ and the IID model in the limit $(1-r)\ll 1$. In the presence of a nonzero 
resetting probability $r$, the positions of the walker are weakly correlated in time and hence it interpolates between the random walk model where the 
positions are strongly correlated and the IID model where the positions are uncorrelated. For arbitrary $r \in [0,1[$, we have 
computed exactly the mean number of records $\langle R_N \rangle$ and shown that it is universal for any $N$, i.e., independent of the jump distribution 
$f(\eta)$, as long as $f(\eta)$ is symmetric and continuous. For example, in the presence of resetting, both random walks with, say a Gaussian jump distribution, as well as L\'evy flights share the same $\langle R_N \rangle$ for
every $N$ and fixed $r$. We also computed the exact universal crossover scaling functions for $\langle R_N \rangle$ in the two limits $r \to 0$ and $r \to 1$. Finally, we have performed numerical simulations which are in excellent agreement with our analytical predictions.

It is well known that for both $r=0$ (the random walk problem without resetting) and $r$ close to $1$ (IID model), not only the mean number of records $\langle R_N \rangle$ 
but also the higher moments of $R_N$, and its full probability distribution, are universal. In this paper, we have seen that the first moment is universal for any given $r \in [0,1[$. 
It is then natural to ask whether the higher moments of $R_N$ are also universal for intermediate values of $r$ between the two known limits $r=0$ and $r$ close to $1$. 
Our preliminary computations and the simulations for the variance seem to indicate that the universality does not extend to higher moments beyond the average, for $0<r<1$. However, one may expect that there will be a vestige of universality still left in the full distribution of $R_N$ at least in the two limiting cases $r$ close to $0$ and $r$ close to $1$. Hence it will be interesting to compute the full distribution of $R_N$ in these two limits -- this is left for future investigations.

\appendix

\section{Computation of $f'_1(u)$}\label{App}

In this Appendix we give the details of the computations of $f_1'(u)$ leading to the expression given in Eq. (\ref{f1prime}) in the main text. 
Our starting point is the expression for the Laplace transform of $f_1(u)$ given in Eq. (\ref{exact_LT_f1}). By performing an integration by parts and using $f_1(0)=0$ from (\ref{f1_explicit}), we find, from 
Eq. (\ref{exact_LT_f1})
\bea \label{LTfprime}
\int_0^\infty f_1'(u) \,e^{-pu} \, du = \sqrt{1+p}\ln\left(\frac{1+p}{p}\right) \;.
\eea
To proceed, we use an integral representation of $\ln{((1+p)/p)}$ to write the rhs of (\ref{LTfprime}) as
\bea \label{app1}
 \sqrt{1+p} \, \ln\left(\frac{1+p}{p}\right) = \int_0^1 \frac{\sqrt{1+p}}{z+p} \, dz \;.
\eea 
The inverse Laplace transform of the integrand in (\ref{app1}) can then be explicitly computed, yielding
\bea \label{app2}
{\cal L}^{-1}_{p\to u} \left(\frac{\sqrt{1+p}}{z+p}\right)  = \frac{e^{-u}}{\sqrt{\pi u}}  + e^{-z\,u} \sqrt{1-z} \; {\rm erf}\left(\sqrt{u(1-z)}\right) \;.
\eea
We now insert this result in the integral over $z$ in Eq. (\ref{app1}) to obtain, using Eq. (\ref{LTfprime})
\bea \label{app3}
f_1'(u) = \int_0^1 {\cal L}^{-1}_{p\to u} \left(\frac{\sqrt{1+p}}{z+p}\right)  \, dz = \int_0^1 \left[ \frac{e^{-u}}{\sqrt{\pi u}}  + e^{-z\,u} \sqrt{1-z} \; {\rm erf}\left(\sqrt{u(1-z)}\right) \right] \, dz \;.
\eea
The integral of the first term in Eq. (\ref{app3}) is trivial. For the second one, we perform the change of variable $x = 1-z$ and use the identity (obtained using Mathematica)
\bea \label{app4}
\hspace*{-0.5cm}\int_0^1 e^{-z\,u} \sqrt{1-z} \; {\rm erf}\left(\sqrt{u(1-z)}\right) \, dz &=& \int_0^1 e^{-u(1-x)}\,\sqrt{x}\, {\rm erf}(\sqrt{u \,x}) \, dx \nonumber \\
&=& e^{-u} \left[ -\frac{1}{\sqrt{\pi\,u}} + \frac{1}{\sqrt{\pi\,u}} \,_2F_2\left(\{1,1\}, \{\frac{1}{2}, 2\},u \right) \right].
\eea
Finally, inserting this result (\ref{app4}) in Eq. (\ref{app3}), yields the result given in Eq. (\ref{f1prime}) in the main text.

%
%%%%%%%%%%%%%%%%%%%%

%

\begin{thebibliography}{1}
%

\bibitem{Cha1952}
Chandler K N 1952 {\it J. Roy. Statist. Soc. Ser. B} {\bf 14} 220
%, The Distribution and Frequency of Record Values,  {\bf 14}, 220 (1952).

\bibitem{FS54}
Foster F G and Stuart A 1954 {\it J. Roy. Statist. Soc.} {\bf 16} 1

\bibitem{hoyt}
Hoyt D V 1981 {\it Climatic Change} {\bf 3} 243
%Weather records and climatic change, {Climatic Change} {\bf 3}, 243 (1981).

\bibitem{basset}
Basset G W 1992 {\it Climatic Change} {\bf 21} 303
%, Breaking recent global temperature records, {Climatic Change} {\bf 21}, 303 (1992).


\bibitem{SZ1999}
Schmittmann B and Zia R K 1999 {\it Am. J. Phys.} {\bf 67} 1269 
%``Weather'' records: Musings on cold days after a long hot Indian summer, Am. J. Phys. {\bf 67}, 1269 (1999).



\bibitem{benestad}
Benestad R E 2003 {\it Climate Res.} {\bf 25} 1
%How often can we expect a record event?, {Climate Res.} {\bf 25}, 1 (2003).

\bibitem{RP2006}
Redner S and Petersen M R 2006 {\it Phys. Rev. E} {\bf 74} 061114
%Role of global warming on the statistics of record-breaking temperatures, {Phys. Rev. E} {\bf 74}, 061114 (2006).

\bibitem{K2007}
Krug J 2007 {\it J. Stat. Mech.}, 07001
%Records in a changing world

\bibitem{WK2010}
Wergen G and Krug J 2010 {\it Europhys. Lett.} {\bf 92} 30008
%Record-breaking temperatures reveal a warming climate, {Europhys. Lett.} {\bf 92}, 30008 (2010).

\bibitem{AB2010}
Anderson A and Kostinski A 2010 {\it J. Appl. Meteor. Clim.} {\bf 49} {1681} 
%Reversible Record Breaking and Variability: Temperature Distributions across the Globe, {J. Appl. Meteor. Clim.} {\bf 49}, {1681} (2010). 





\bibitem{WHK2013}
Wergen G, Hense A and Krug J 2013 {\it Clim. Dynam.} {\bf 22} 1 
%Record occurrence and record values in daily and monthly temperatures, Clim. Dynam. {\bf 22}, 1 (2013).



\bibitem{records_finance}
Barlevy G and Nagaraja H N 2006 {\it J. Appl. Prob.} {\bf 43}, 1119



\bibitem{WBK2011}
Wergen G, Bogner M and Krug J 2011 {\it Phys. Rev. E} {\bf 83} 051109 
%, Record statistics for biased random walks, with an application to financial data, {Phys. Rev. E} {\bf 83}, 051109 (2011). 

\bibitem{SL2014}
Sabir B and Lanthanum M S 2014 {\it Phys. Rev. E} {\bf 90} 032126  
%Record statistics of financial time series and geometric random walks, Phys. Rev. E {\bf 90}, 032126 (2014).


\bibitem{records_hydrology}
Matalas N C 1997  {\it Climatic Change} {\bf 37} 89 
%Stochastic Hydrology in the Context of Climate Change, {Climatic Change} {\bf 37}, 89 (1997)



\bibitem{Gembris}
Gembris D, Taylor J G and Suter D 2002 {\it Nature} {\bf 417} 506
%Sports statistics: Trends and random fluctuations in athletics, {Nature} {\bf 417}, 506 (2002). 

\bibitem{sports}
Ben-Naim E, Redner S and Vazquez F 2007 {\it Europhys. Lett.} {\bf 77}, 30005
%Scaling in Tournaments, {Europhys. Lett.} {\bf 77}, 30005 (2007).


\bibitem{FWK2012}
Franke J, Wergen G and Krug J 2012 {\it Phys. Rev. Lett.} {\bf 108}, 064101
%Correlations of record events as a test for heavy-tailed distributions, 
%Phys. Rev. Lett. {\bf 108}, 064101 (2012).

\bibitem{GL2008}
Godr\` eche C and Luck J M 2008 {\it J. Stat. Mech.} P11006

\bibitem{sibani}
Sibani P, Rodriguez G F and Kenning G G 2006 {\it Phys. Rev. B} {\bf 74} 224407 
%Intermittent quakes and record dynamics in the thermoremanent magnetisation of a spin-glass, {Phys. Rev. B} {\bf 74} 224407 (2006); P. Sibani, Linear response in aging glassy systems, intermittency and the Poisson statistics of record fluctuations, {Eur. Phy. J. B} {\bf 58}, 483 (2007).

\bibitem{MBK2019}
Majumdar S N, von Bomhard Ph and Krug J 2019 {\it Phys. Rev. Lett.} {\bf 122} 158702


\bibitem{MPS2020}
Majumdar S N, Pal A and Schehr G 2020 {\it Phys. Rep.} {\bf 840} 1 (2020).
%Extreme value statistics of correlated random variables: a pedagogical review



\bibitem{review_wergen}
Wergen G 2013 {\it J. Phys. A: Math. Theor.}, {\bf 46}, 223001 
%Records in stochastic processesÑtheory and applications.

\bibitem{review_records}
Godr{\` e}che C, Majumdar S N and Schehr G 2017 {\it J. Phys. A: Math. Theor.} {\bf 50} 333001
 %{\it Record statistics of a strongly correlated time series: random walks and L{\'e}vy flights}, J. Phys. A: Math. Theor. 50, 333001 (2017).


\bibitem{Res1987}
Resnick S I 1987 {\it Extreme Values, Regular Variation, and Point Processes} (Springer: New York) 

\bibitem{ABN1992}
Arnold B C, Balakrishnan N and Nagaraja H N 1998 {\it Records} (New York: Wiley)

\bibitem{BG2001}
Bunge J and Goldie C M 2001 {\it Handbook of Statistics} {\bf 19} 277
%Record sequences and their applications, Handbook of Statistics {\bf 19}, 277 (2001).

\bibitem{Nevzorov}
Nevzorov V B 2004 {\it Records: Mathematical Theory} (Providence, RI: American Mathematical Society) 

\bibitem{Feller} Feller W 1966  {\it Introduction to Probability Theory and Its Applications} Vol. 2 (Wiley: New York)


\bibitem{KJ2005}
Krug J and Jain K 2005, {\it Physica A}, {\bf 358} 1
%Breaking records in the evolutionary race

\bibitem{FKAK2011}
Franke J, Kl{\"o}zer A, Arjan G J and Krug J 2011, {\it PLos Comput. Biol.} {\bf 7} e1002134

\bibitem{PSNK2015}
Park S C, Szendro I G, Neidhart J and Krug J 2015, {\it Phys. Rev.} E {\bf 91}, 042707 

\bibitem{PNK2016}
Park S C, Neidhart J and Krug J 2016, {\it J. Theor. Bio.} {\bf 397} , 89

\bibitem{PK2016}
Park S C and Krug J 2016, {\it J. Phys. A: Math. Theor.} {\bf 49}, 315601














\bibitem{MZ2008}
Majumdar S N and Ziff R M 2008 {\it Phys. Rev. Lett.} {\bf 101} 050601
%Universal record statistics of random walks and L\'evy flights, {Phys. Rev. Lett.} {\bf 101}, 050601 (2008). 



\bibitem{PLDW2009}
Le Doussal P and Wiese K J 2009  {\it Phys. Rev. E} {\bf 79} 051105
%Driven particle in a random landscape: disorder correlator, avalanche distribution and extreme value statistics of records, Phys. Rev. E {\bf 79}, 051105 (2009).


\bibitem{Satya2010}
Majumdar S N 2010 {\it Physica A} {\bf 389} 4299
%Leuven's lectures

\bibitem{Sanjib2011} 
Sabhapandit S 2011 {\it Europhys. Lett.} {\bf 94} 20003
%Record Statistics of Continuous Time Random Walk, Europhys. Lett. {\bf 94}, 20003 
%(2011).



\bibitem{MSW2012}
Majumdar S N, Schehr G and Wergen G 2012 {\it J. Phys. A: Math. Theor.} {\bf 45} 355002 
 

\bibitem{WMS2012}
Wergen G, Majumdar S N and Schehr G 2012 {\it Phys. Rev. E} {\bf 86} 011119
% Record statistics for multiple random walks, {Phys. Rev. E} {\bf 86}, 011119 (2012). 

\bibitem{EKMB2013}
Edery Y, Kostinski A B, Majumdar S N and Berkowitz B 2013, {\it Phys. Rev. Lett.} {\bf 110}, 180602
%Record-breaking statistics for random walks in the presence of measurement error and noise

\bibitem{GMS2015b}
Godr\`eche C, Majumdar S N and Schehr G 2015 {\it J. Stat. Mech.} P07026 
%, {Record statistics for random walk bridges}, J. Stat. Mech. P07026 (2015).

\bibitem{GMS2016}
Godr\`eche C, Majumdar S N and Schehr G 2016 {\it Phys. Rev. Lett.} {\bf 117} 010601
%Exact statistics of record increments of random walks and L\'evy flights, Phys. Rev. Lett. {\bf 117}, 010601 (2016).


\bibitem{Cha15b}
Challet D 2017 {\it Appl. Math. Fin.} {\bf 24} 1
%Sharper asset ranking from total drawdown durations. Applied Mathematical Finance, 24(1), 1-22.



\bibitem{MMS2020}
Mounaix Ph, Majumdar S N and Schehr G 2020 {\it J. Phys. A: Math. Theor.} {\bf 53} 415003
%Statistics of the Number of Records for Random Walks and LŽvy Flights on a 1D Lattice


\bibitem{MDMS2020a}
Mori F, Le Doussal P, Majumdar S N and Schehr 2020 {\it Phys. Rev. Lett.} {\bf 124} 090603
%Universal survival probability for a d-dimensional run-and-tumble particle, Phys. Rev. Lett. 124, 090603 (2020).

\bibitem{MDMS2020b}
Mori F, Le Doussal P, Majumdar S N and Schehr G 2020 {\it Phys. Rev. E} {\bf 102} 042133
%Universal Properties of a Run-and-Tumble Particle in Arbitrary Dimension, Phys. Rev. E 102, 042133 (2020)

\bibitem{LM2020}
Lacroix-A-Chez-Toine B, and Mori F 2020 {\it J. Phys. A} {\bf 9} 101
%Universal survival probability for a correlated random walk and applications to records. Journal of Physics A: Mathematical and Theoretical, 53(49), 495002.

\bibitem{Kea2020}
Kearney M J 2020, {\it J. Stat. Mech.} 023206 

\bibitem{MMS2021}
Majumdar S N, Mounaix Ph and Schehr G 2021, {\it J. Phys. A: Math. Theor.} {\bf 54}, 315002
%Universal record statistics for random walks and LŽvy flights with a nonzero staying probability




\bibitem{GL2021}
Godr\`eche C and Luck J M 2021 {\it preprint arXiv:2109.05582}

\bibitem{MMR2021}
Maria Schimmenti V, Majumdar S N and Rosso A 2021 {\it preprint arXiv:2106.01411}
%Statistical properties of avalanches via the c-record process




\bibitem{EM1} Evans M R and Majumdar S N 2011
% Diffusion with stochastic resetting,
 {\it  Phys. Rev. Lett.}  {\bf 106}, 160601 


\bibitem{EM2} Evans M R and Majumdar S N 2011 
% Diffusion with optimal resetting,
 {\it  J. Phys. A: Math. Theor.}  {\bf 44}, 435001 


\bibitem{MV13} Montero M and  Villarroel J 2013 
%Monotonous continuous-time random walks with drift and stochastic reset events,
{\it Phys. Rev. E} {\bf 87}, 012116 


\bibitem{EM14}
Evans M R and  Majumdar S N 2014
%Diffusion with resetting in arbitrary spatial dimension
{\it  J. Phys. A: Math. Theor.} {\bf 47},  285001  

\bibitem{KMSS14}
Ku\'smierz L,  Majumdar S N,  Sabhapandit S and  Schehr G 2014
%First Order Transition for the Optimal Search Time of L\'evy Flights with Resetting
{\it Phys. Rev. Lett.} {\bf 113}, 220602 


\bibitem{MSS15a}
 Majumdar S N,  Sabhapandit S and  Schehr G 2015
%Dynamical transition in the temporal relaxation of stochastic processes under resetting
{\it Phys. Rev. E} {\bf 91}, 052131 

\bibitem{MSS15b}
 Majumdar S N,  Sabhapandit S and  Schehr G 2015
%Random walk with random resetting to the maximum position
{\it Phys. Rev. E} {\bf 92}, 052126 


\bibitem{Pal15}
Pal A 2015
%Diffusion in a potential landscape with stochastic resetting
{\it Phys. Rev. E} {\bf 91} 012113



\bibitem{KGN15}
Ku\'smierz L and  Gudowska-Nowak E 2015
%Optimal first-arrival times in L\'evy flights with resetting
{\it Phys. Rev. E}  {\bf 92} 052127



\bibitem{PKE16}
Pal A, Kundu A and  Evans M R 2016,
%Diffusion under time-dependent  resetting
{\it  J. Phys. A: Math. Theor.} {\bf 49},  225001  


\bibitem{NG16}
Nagar A and  Gupta S 2016
%Diffusion with stochastic resetting at power-law times
{\it  Phys. Rev. E}  {\bf 93},  060102 (R)  

\bibitem{Reuveni16}
Reuveni S 2016
%Optimal Stochastic Restart Renders Fluctuations in First Passage Times Universal
{\it Phys. Rev. Lett.}  {\bf 116}, 170601

\bibitem{PR17}
Pal A and Reuveni S 2017
%First Passage under Restart
{\it Phys. Rev. Lett.} {\bf 118}, 030603




\bibitem{CS18}
Chechkin A, Sokolov I M 2018
%Random Search with Resetting: A Unified Renewal Approach,
{\it Phys. Rev. Lett.} {\bf 121}, 050601

\bibitem{EM18}
Evans M R  and Majumdar  S N 2018
%Run and tumble particle under resetting: a renewal approach
{\it  J. Phys. A: Math. Theor.} {\bf 51} 475003


\bibitem{MPCM19a}
Mas\'o-Puigdellosas A, Campos D, and M\'endez V 2019
%Transport properties and first-arrival statistics of random motion with stochastic reset times
{\it Phys. Rev. E}  {\bf 99}, 012141 

\bibitem{MMSS21}
Majumdar S N, Mori F, Schawe H and Schehr G 2021 {\it Phys. Rev. E} {\bf 103}, 022135
%Mean perimeter and area of the convex hull of a planar Brownian motion in the presence of resetting



\bibitem{reset_review}
Evans M R, Majumdar S N and Schehr G 2020 {\it J. Phys. A: Math. Theor.} {\bf 53}, 193001 


\bibitem{MO18}
Majumdar S N and Oshanin G 2018
%Spectral content of fractional Brownian motion with stochastic reset
{\it J. Phys. A: Math. Theor} {\bf 51}, 435001

\bibitem{SSKP21}
Stojkoski V, Sandev T, Kocarev L and Pal A 2021 {\it preprint arXiv:2107.11686}
%Autocorrelation functions and ergodicity in diffusion with stochastic resetting
%optimal

\bibitem{BBPMC20}
Besga B, Bovon A, Petrosyan A, Majumdar S N and Ciliberto S 2020 {\it Phys. Rev. Res.} {\bf 2}, 032029
%Optimal mean first-passage time for a Brownian searcher subjected to resetting: experimental and theoretical results

\bibitem{TPSRR20}
Tal-Friedman O, Pal A, Sekhon A, Reuveni S and Roichman Y 2020 {\it J. Phys. Chem. Lett.} {\bf 11}, 7350 
%Experimental realization of diffusion with stochastic resetting


\bibitem{FBPCM21}
Faisant F, Besga B, Petrosyan A, Ciliberto S and Majumdar S N 2021 {\it preprint arXiv:2106.09113}
%Optimal mean first-passage time of a Brownian searcher with resetting in one and two dimensions: Experiments, theory and numerical tests



\bibitem{EMM13}  Evans M R,   Majumdar S N and  Mallick K 2013
%Optimal diffusive search: nonequilibrium resetting versus
%equilibrium dynamics,
{\it J. Phys. A: Math. Theor.} {\bf 46}, 185001 


\bibitem{CS15}
Christou C and  Schadschneider A 2015
%Diffusion with resetting in bounded domains
{\it J. Phys. A: Math. Theor.}  {\bf 48},  285003

\bibitem{BBR16}
Bhat U, De Bacco C and Redner S 2016 {\it J. Stat. Mech} 083401



\bibitem{BRR20}
De Bruyne B, Randon-Furling J and Redner S 2020 {\it Phys. Rev. Lett.} {\bf 125}, 050602
%Optimization in First-Passage Resetting




\bibitem{Bres2020}
Bressloff P C 2020 {\it J. Phys. A: Math. Theor.} {\bf 53}, 275003 
%Switching diffusions and stochastic resetting

\bibitem{Bres2021}
Bressloff P C 2021, J. Stat. Mech. 063206 
%Drift-diffusion on a Cayley tree with stochastic resetting: the localization-delocalization transition



\bibitem{Poll52}
Pollaczek F 1952 {\it C. R.} {\bf 234}, 2334

\bibitem{SA1954}
Sparre Andersen E 1954 {\it Math. Scand.} {\bf 2} 195


\bibitem{Spi56}
Spitzer F 1956 {\it Trans. Am. Math. Soc.} {\bf 82}, 323




\bibitem{Iva94}
Ivanov V V 1994 {\it Astron. Astrophys.} {\bf 286}, 328






\bibitem{CM2005}
Comtet A and Majumdar S N 2005 {\it J. Stat. Mech.} 06013
%Precise Asymptotics for a Random Walker's Maximum




\bibitem{MMS2017}
Majumdar S N, Mounaix Ph. and Schehr G 2017, {\it J. Phys. A: Math. Theor.} {\bf 50}, 465002
%Survival Probability of Random Walks and LŽvy Flights on a Semi-Infinite Line


\bibitem{MMS2014}
Majumdar S N, Mounaix Ph and Schehr G 2014 {\it J. Stat. Mech}, 09013
%On the Gap and Time Interval between the First Two Maxima of Long Random Walks








\end{thebibliography}
\end{document}